\documentclass[conference]{IEEEtran}

\usepackage{tikz}
\usepackage{standalone}

\usepackage[normalem]{ulem} 
\usepackage{url}
\usepackage{cite}
\usepackage{amsmath,amssymb,amsfonts}
\usepackage{algorithmic}
\usepackage{graphicx}
\usepackage{textcomp}
\usepackage{xcolor}
\def\BibTeX{{\rm B\kern-.05em{\sc i\kern-.025em b}\kern-.08em
    T\kern-.1667em\lower.7ex\hbox{E}\kern-.125emX}}
\begin{document}

\title{Agent-Based Modelling of Ethereum Consensus}

\author{\IEEEauthorblockN{Benjamin Kraner\IEEEauthorrefmark{1}\IEEEauthorrefmark{4}, 
Nicol\`{o} Vallarano\IEEEauthorrefmark{1}\IEEEauthorrefmark{2},
Claudio J. Tessone\IEEEauthorrefmark{1}\IEEEauthorrefmark{2} and Caspar Schwarz-Schilling\IEEEauthorrefmark{3}
}%
\thanks{\copyright 2023 IEEE.  Personal use of this material is permitted. Permission from IEEE must be obtained for all other uses, in any current or future media, including reprinting/republishing this material for advertising or promotional purposes, creating new collective works, for resale or redistribution to servers or lists, or reuse of any copyrighted component of this work in other works.} \\

\thanks{The work was supported by a grant from the Ethereum Foundation as part of the Academic Grants Round.}
\IEEEauthorblockA{\IEEEauthorrefmark{1}Blockchain \& Distributed Ledger Technologies Group,\\Department of Informatics,  University of Zurich, Switzerland}
\IEEEauthorblockA{\IEEEauthorrefmark{2}UZH Blockchain Center,  University of Zurich, Switzerland}
\IEEEauthorblockA{\IEEEauthorrefmark{3}Ethereum Foundation}
\IEEEauthorblockA{\IEEEauthorrefmark{4} Email: benjamin.kraner@uzh.ch}
}

\IEEEoverridecommandlockouts


\maketitle
\IEEEpubidadjcol

\begin{abstract}

This paper presents a study of the Poof-of-Stake (PoW) Ethereum consensus protocol, following the recent switch from Proof-of-Work (PoS) to Proof-of-Stake within Merge upgrade. The new protocol has resulted in reduced energy consumption and a shift in economic incentives, but it has also introduced new threat sources such as chain reorganizations and balancing attacks. Using a simple and flexible agent-based model, this study employs a time-continuous simulation algorithm to analyze the evolution of the blocktree and assess the impact of initial conditions on consensus quality. The model simulates validator node behavior and the information propagation throughout the peer-to-peer network of validators to analyze the resulting blockchain structure. Key variables in the model include the topology of the peer-to-peer network and average block and attestation latencies. Metrics to evaluate consensus quality are established, and means to observe the model's responsiveness to changes in parameters are provided. The simulations reveal a phase transition in which the system switches from a consensus state to a non-consensus state, with a theoretical justification presented for this observation.

\end{abstract}

\section{Introduction}
Ethereum has recently undergone some major upgrades. The most recent (and arguably one of the most important in Ethereum's history) of these upgrades was the Merge,\footnote{https://ethereum.org/en/upgrades/merge/} executed on September 15, 2022, with the Paris upgrade. The Merge marked the passage of Ethereum from the original Proof-of-Work consensus protocol \cite{Tasca_Tessone_2019,tree} to the actual Ethereum network using Proof-of-Stake as a consensus protocol. While Ethereum upgrades are yet to be finished, e.g. the eventual introduction of sharding \cite{bez2019scalability}, the switch to Proof-of-Stake consensus was a revolution for Ethereum: it immensely reduced the energy consumption of Ethereum, it disrupted the economic incentives structure by replacing miners with validators and it set the course of Ethereum's progress further from the original Bitcoin paradigm \cite{nakamoto2008bitcoin}.
As a reference model of the Proof-of-Stake (PoS) Ethereum consensus protocol we will consider the \emph{Gasper} protocol \cite{buterin2020combining}, a combination of variants of Casper FFG (Friendly Finality Gadget)\cite{buterin2017casper} and LMD GHOST (Latest Message Driven Greediest Heaviest Observed Subtree) \cite{sompolinsky2015secure} which was presented as an idealized abstraction of the proposed Ethereum implementation \cite{buterin2020combining}.

While the switch to Proof-of-Stake indeed brought many advantages, it also led to novel threat sources. 
In preparation to the Merge many researchers in recent years have focused on analysing attack strategies on Gasper and devising possible counter-measures and protocol improvements \cite{d2022no, daian2019flash}. Two main examples of attacks in the literature are chain reorganisations \cite{neuder2021low} and balancing attacks\cite{neu2021ebb,neuder2021low}. 
To execute chain reorganisations (also known as \emph{reorgs}) attackers have to withhold blocks and attestations in order to censor blocks of honest participants. Balancing attacks\cite{neu2021ebb,neuder2021low} are attacks in which an attacker strategically times information release, so as to split the view of honest validators, with the goal of increasing time to block finality. Further possible combinations and modifications of the two exist \cite{schwarz-schilling2021attacks}.

The present work offers a contribution to the field by providing a simple and flexible agent-based model of PoS Ethereum consensus. The underlying agent-based simulation algorithm is based on the family of Gillespie models \cite{gillespie1976general} which allow for continuous time simulation, increasing the simulation efficiency. We follow the steps of previous work in the field \cite{tessone2021stoc} mainly related to PoW: the minimalist approach allowed the simulation of strategic deviations from the honest protocol, as analyzed in \cite{CSS2021,CSS2022}, and recently identified in real data \cite{Li2020,Li2020b}, and even agent behaviour in absence of rewards, as seen in  \cite{Kraner2022}.
Our model simulates, under different initial conditions, the evolution of the \emph{blocktree}, the network of all blocks produced, where each block is linked to its previous block, including blocks which are not part of the canonical chain. 
The two main parameters of the model are diffusion latencies for the two different gossip events that may happen between Ethereum peers: a block gossip, the event where peers exchange information about the blocks they have received, and an attestation gossip, the event where peers communicate to each other their opinion on the blocks validity, called an attestation. We will describe what attestations are in detail in section \ref{sec:system}.
By modelling the exchange of information between peers, we reconstruct the consensus evolution and eventually analyse how the consensus quality is affected by the initial conditions, comprising the peer-to-peer underlying topology and the gossip diffusion latency. Our modelling approach is time-continuous, in contrast to others \cite{fadda2022consensus} that are based on discrete time.

The analysis of consensus quality is conducted by studying the topological properties of the blocktree resulting from a model simulation, eventually averaged over multiple realisations in order to achieve statistical significance.

In addition to the present paper, we also published an open source github repository containing the code necessary to interact with the agent-based consensus model and replicate our results.\footnote{https://github.com/benckj/ethereum-consensus-abm}


The paper is organized as follows: in section \ref{sec:system} we briefly describe the Ethereum Beaconchain Proof-of-Stake protocol for the non familiar readers. In section \ref{sec:model} we proceed to present our modelling of the protocol. In section \ref{sec:results} we present and test measures to asses consensus quality under different initial conditions. Finally, in section \ref{sec:conclusion} we discuss the model performance and highlight future extensions.

\section{The System}\label{sec:system} 
The Ethereum network is a decentralized, peer-to-peer network. It consists of nodes, which are computers (servers) that maintain a copy of the blockchain. Nodes can offer different functionality depending on the users preferences, e.g. light nodes vs full nodes. Some nodes, incentivized by reward mechanisms, fulfill a special role; they act as validators. 

\subsection{System Outline}
In the Proof-of-Stake Ethereum consensus, validators are participants involved in the consensus process. As suggested by the consensus algorithm's name, validators are required to lock up stake as collateral. To start, a 32 Ether (ETH) deposit must be submitted to the deposit contract,\footnote{https://github.com/ethereum/consensus-specs/blob/dev/specs/phase0/deposit-contract.md} which maintains a Merkle root for all deposits. A validator serves as the basic unit of consensus engagement, allowing for the activation of multiple validators to increase the total stake. After depositing, a validator is activated following a minimum of 4 epochs.\footnote{https://github.com/ethereum/consensus-specs/blob/dev/specs/phase0/beacon-chain.md\#time-parameters} In case several validators are in line for activation, a queue forms. Upon activation, validators are given responsibilities such as block proposal and attestation; for additional details, refer to section \ref{sub:consensus}. To perform their duties, validators must run a node, which facilitates communication and message exchange within the gossip network.\footnote{https://github.com/ethereum/consensus-specs/blob/dev/specs/phase0/p2p-interface.md} Validators receive ETH-based rewards as incentives for proper conduct, compensating for the associated costs.

Validators independently establish the canonical chain to accurately propose or attest to blocks. This is achieved by adhering to the fork choice rule, a mechanism that identifies the head of the chain based on the local view of the block tree and received attestations. The fork choice rule is referred to as the Latest Message Driven Greediest Heaviest Observed SubTree (LMD GHOST).

In the LMD framework, validators maintain a table of the 'latest message' (where message = a vote for a block deemed as the head of the chain) received from all other validators. When a valid vote is received from a validator, the 'latest message' table entry for that validator is updated only if the new vote comes from a slot strictly later than the existing 'latest message' table entry. As a result, if a validator encounters two equivocating votes from the same validator for the same slot, the validator considers the vote that arrived first.

Initiating the fork choice rule from the most recent justified checkpoint, it advances down the block tree. In the GHOST protocol \cite{sompolinsky2015secure}, if a fork arises, the heaviest subtree is chosen. The term 'heavier' pertains to the concept of weight, which measures the total ETH attesting to a given block or its descendants, considering only the latest attestations from active validators.

\subsection{Consensus Outline} \label{sub:consensus} 
The consensus mechanism of Ethereum is a composition of the LMD GHOST fork choice rule as its foundation, accompanied by the Casper FFG as a finality mechanism, working together through two separate phases and distinct time scales.

In the first phase, LMD GHOST operates on a smaller time scale, with time segments referred to as slots, each lasting $2\Delta$. A \emph{block proposer} and a \emph{committee} of $W$ validators are randomly chosen from the pool of $N$ validators during each slot. The LMD GHOST rule is applied to determine a canonical block and its preceding chain in a node's view at slot $t$: "Beginning with the highest block $b_0$ 'justified' by Casper FFG (explained later), calculate the sum of unique valid votes (i.e., only from earlier than the current slot, and not voting on future blocks) for each child block $b$ and its descendants; count only the latest vote (LMD) cast by each validator. Select the child block $b^*$ with the highest weight (GHOST) (breaking ties adversarially). Iterate ($b_0 \leftarrow b^*$) until a leaf block is encountered. Output that leaf block." At the start of each slot, the proposer selects a block using LMD GHOST and adds a new proposal to it. Committee members of a slot vote for a block using LMD GHOST in their local view under two conditions: (a) they receive a valid block from the expected block proposer for the assigned slot, or (b) 4 seconds of the slot have passed, whichever happens first. These votes are also known as \emph{attestations}. 

While in the remainder of this paper we are mostly concerned with LMD GHOST, for completeness, we provide a brief overview of the second phase. It takes place on a larger time scale. Casper FFG operating within epochs composed of $32$ slots. At a high level, Casper FFG is a leaderless, two-phase propose-and-vote-style Byzantine fault-tolerant (BFT) consensus protocol (similar to Chained HotStuff \cite{yin2018hotstuff}), integrated into a blockchain protocol within the chained framework. The LMD GHOST fork choice layer is responsible for generating proposals consistently across honest nodes. Casper FFG functions as follows: Blocks become justified when a super-majority ($2N/3$) votes 'for them', and later become finalized when a super-majority votes 'from them' for a subsequent block. The genesis block is, by definition, both justified and finalized. Validators cast their votes during an epoch for the so-called epoch boundary blocks, which are the remaining leaf blocks after truncating the block tree to only include blocks from the previous epoch. Validators vote for the highest epoch boundary block consistent with the highest justified block they have observed, extending the most recent finalized block they have observed. The super-majority requirement for proposal progression and the two-phase confirmation process (referred to as \emph{finalization}) ensures that Casper FFG remains secure even during temporary network partitions. The confirmation rule at the Casper FFG level involves outputting the most recent finalized block and its prefix.

\section{The model}\label{sec:model}
\subsection{Model settings}

The emergence of the Ethereum canonical blockchain is the result of the independent execution of the Ethereum consensus protocol by independent agents on a peer-to-peer network. We represent the peer-to-peer network as a graph where nodes represent peers and the links represent communication channels. Using a Gillespie model (events-based) the agents (the peers) will interact trough channels on the basis of a collection of events. 
In this subsection we describe the agents and the parameters which completely describe the model state.
 
\paragraph{Nodes} The basic model constituents are \emph{nodes}, intended to represent a peer participating in the peer-to-peer (p2p) communication network which acts a the cornerstone of the Ethereum Beaconchain. Every peer spreads information in the network by gossiping it  in the form of messages (gossiping). As we are studying the consensus mechanisms of Ethereum, we are especially interested in the key contributors: the \emph{validators}. Validators are nodes which stake their ETH in order participate actively in the consensus mechanism. Our model neglects the presence of other (non-validating nodes) nodes, as we are interested in the consensus mechanisms induced by the functioning of the validator nodes. Other (relay) nodes could be either integrated as an extension of the model or integrated into our setting by setting an effective, higher latency parameter.

We consider a static set of nodes (validators). Node churn and node renewal have characteristic time-scales much longer than those of block creation and consensus \cite{tessone2021stoc}, that are the focus of this study.
The static node-set is also consistent with the consensus description from \cite{buterin2020combining}. 
We will introduce these  dynamical elements into the model in the future, to   analyse longer time-scales in the Ethereum consensus.

\paragraph{Topology}
Here we consider the set of links connecting the node-set $N$: a directed link from $i$ to $j$ exists when $ (i,j) \in E$. 
The graph of the peer-to-peer network is completely described by the couple $\mathcal{G}= (N, E)$.

In the experiments showed in section \ref{sec:results} we used an Erdős–Rényi (ER)\cite{erdHos1960evolution} random graph to sample the underlying peer-to-peer graph.
The choice was made because of the ER graphs predictable behaviour\cite{chung2001diameter} in terms of diameter and connectedness. As the interested reader will find out in the attached code, different choices are available and were explored.

We assume that the p2p topology changes at a larger scale than the protocol consensus. Thus, the topology in our model is static, meaning the number and identity of nodes does not change, neither do the links among them.
 This assumption will be relaxed  future works to introduce dynamical effects (node churn, rewiring of the links etc.) to observe longer term effect and increase the models precision.

\paragraph{Block} 
Blocks represent the blocks of the Ethereum Beaconchain which are exchanged between the nodes. A block represented by $b$ contains a link to his parent $p(b)$, enabling them to form a blockchain. They are created (proposed for inclusion in the mainchain) by a \emph{proposer} at the begining of the slot and subsequently broadcast over the peer-to-peer network.

\paragraph{Attestations}
Attestations are published by the validators of a slot. They attest for the block they deem to be the head of the (canonical) chain once per slot (i.e. they are members of the relevant committee). Either they receive the block which is proposed in the current slot, before the attestation threshold is reached and they attest for it. If they do not receive the current slot block, they will attest for the block which they deem is head of the chain (by LMD GHOST) at the time of the attestation threshold.
Attestations in our model are equipped with a pointer to the agent that issued it, a pointer to the block it is attesting for and a number indicating the slot they were issued in (in order to allow nodes to check which attestations are the newest).

\paragraph{Stake}
The agents of the model, the nodes of the network, represent the validators of the Ethereum network. As described in section \ref{sec:system}, each validator node (in order to be activated) has to stake 32 Ether. 
In the model we will assume each node to represent a single validator instance with full stake. While we have homogeneity in stake distribution at node level, heterogeneity emerges in attestations distribution (each attestation is weighted by stake) per block. Future work might change the stake of nodes or give an entity control over multiple nodes, to represent the stake differentials encountered in Ethereum.


\subsection{Model simulation}
The evolution of the model is dictated by events, which change the states of agents and therefore the state of the system. We distinguish five different events in our model, which can be categorized in deterministic and stochastic events. The deterministic events consist of \textbf{i.} slot boundary, \textbf{ii.} epoch boundary and \textbf{iii.} attestation threshold events. The stochastic events dictate the exchange of information/data over the peer-to-peer network and consist of \textbf{iv.} block gossip and \textbf{v.} attestation gossip events.

\subsubsection{Deterministic Events}
The deterministic events replicate the structure of the Ethereum protocol, and enable us to structure time into slots and epochs underlying the consensus mechanisms of Ethereum.

\paragraph{Slot Boundary}
As in Ethereum each slot lasts for $12$ seconds, upon activation a \emph{proposer} is selected randomly from the set of validators and proposes a block. A committee of validators (sampled during the epoch boundary event) is activated to attest for the current slot. 

\paragraph{Epoch Boundary}
Derived from the slot boundary event, the epoch boundary event happens before every $m$ slot boundary events. It executes before the slot boundary event, but happens at the same (model) time as subsequent slot boundary event. The epoch boundary event samples the committees for the following $m$ slots. It selects $\lfloor V/m \rfloor$ for each committee and distributes each remaining validator to a distinct committee.
In the actual Ethereum implementation $m=32$. In our simulations we set $m=1$ in order to simulate local consensus of a broader group of validators. Since we are not interested in the finality of blocks but rather the consensus on a smaller timescale.

\paragraph{Attestation threshold}
The attestation threshold event triggers all committee members (i.e. validators from the current slot) that did not yet issue an attestation for the current slot to issue an attestation on their current head of the chain. It is scheduled such that it happens $4$ seconds into the slot (after the slot boundary event). This mimics the behaviour of the Ethereum protocol: if the block has not arrived 4 seconds in the slot, then (honest) validators attest the previous head block. As block diffusion through the network is not immediate, the latency and the specific path of a block, may lead to some nodes not receiving the proposed block on time.

\subsubsection{Stochastic Events}
The stochastic events dictated the flow of data and information on the peer-to-peer network. They model the latency inherent to the gossiping between nodes. We distinguish between the block and attestation gossip as they may differ in size and ceteris paribus the time it takes for them to travel the network can be modeled accordingly. The underlying assumptions for both stochastic events are the same: we assume that the number of blocks (attestations) follows a Poisson distribution with parameter $\tau$ and therefore the  inter-arrival time between two blocks (attestations) follows an exponential distribution with parameter $-\tau^{-1}$. Effectively we are modelling the time it takes for a block to be sent after its arrival at node A to node B to be distributed according to $\hbox{exp}(-\tau^{-1})$. As this processes happen on every edge of the network and in a bidirectional way, we can distinguish $2E$ independent (Poisson) processes, and thus for each event the system-wide inter-arrival time will be given by $\hbox{exp}(-(2E\tau)^{-1})$.

\paragraph{Block gossip}
With parameter $\tau_{\text{block}}$, representing the latency of blocks, blocks are exchanged between edges. 
We select a (directed) edge of the network at random, which defines a sender and a receiver. The receiver will receive the complete canonical chain (genesis to local head block) from the sender. Cached attestations will be matched against the newly arrived blocks and upon a match added to the set of \emph{active} attestations. 

\paragraph{Attestations gossip}
With parameter $\tau_{\text{attestation}}$, representing the latency of attestations, attestations are exchanged between edges. 
We select a (directed) edge of the network at random, which defines a sender and a receiver. From the set of (latest) attestations of the sender, an attestation (which has not yet been relayed to the receiver) will be selected randomly and copied to the receivers set of attestations. Should the block of the attestation be unknown, the attestation is cached, and compared upon the arrival of new blocks. 

\begin{figure}[!t]
\centerline{
\includegraphics[width=\columnwidth]{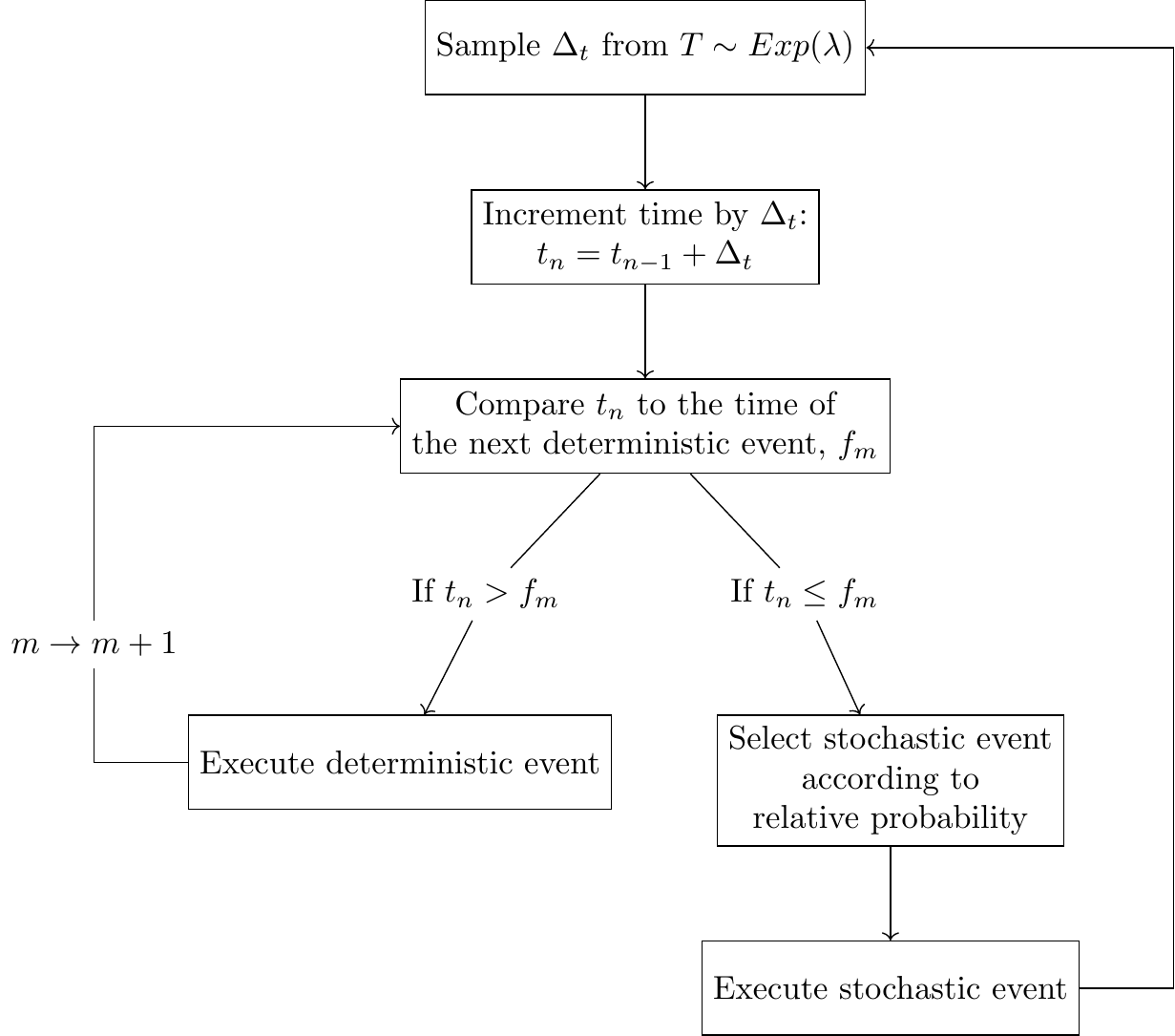}}
\caption{Agent-based model event algorithm flow at a generic point in time $t_{n-1}$. The algorithm is described in full detail in subsection \ref{subsec:algo}.}
\label{fig:workflow}
\end{figure}

\subsubsection{The model algorithm}\label{subsec:algo}
\paragraph{Evolution}
The model can be efficiently simulated using a variation of Gillespie algorithm \cite{gillespie1976general,tessone2021stoc}. 
We consider the p2p network previously introduced, where nodes are peers and links are channels of communication for peers. At time $t$, each peer $i$ contains a local copy of the blockchain $C_i (t)$, commonly initialised at time $t = 0$ with a single, genesis block. In addition to this, validators node also keep a record of all latest attestations by other validators. This can be imagined as a dictionary where the keys are node ids and the item is the latest block attested. The attestation record is initialised as an empty dictionary and populated step by step.

The model we use here is a modification of the original Gillespie algorithm; in order to account for the presence of fixed-time events we had to rewrite the process, by dropping the dynamic evolution of the events' types probability parameter.

We have two kind of stochastic events in the system: attestation gossip events and block gossip events. To each kind we associate a parameter $\tau_{\text{attestation}}$ and $\tau_{\text{block}}$. This parameter can be interpreted as the expected waiting time between two events (i.e. $\tau_{\text{block}}$ is the average waiting time between two block gossiping events). The underlying assumption the Gillespie model take is that the waiting time between two stochastic events of the same type $T_{e}$ for $e\in\{ \text{block}, \text{attestation}\}$ follows an exponential distribution of parameter $\lambda_{e} = \tau^{-1}_{e}$.

We denote the waiting time before the next stochastic event (be it a block or attestation event) as $T \approx min(T_\text{block}, T_\text{attestation})$. It follows that $T$ is distributed as an exponential variable of parameter $\lambda = 2E(\lambda_{\text{block}} + \lambda_{\text{attestation}})$.

If there only were block and attestation gossip events in the model that would be enough to set up the Gillespie algorithm but in our case we also have fixed-time events in the events set we should account for. 
To do so we order all fixed-time events $(f_0, \dots, f_m)$, where $m$ is the total number of fixed events, by their execution time $f_i$.

At the beginning of the simulation $t_0$ we sample the waiting time of the next stochastic event $\Delta_t$ and we compare $t_0+\Delta_t$ with the first fixed-time event happening time $f_0$. If $t_0+\Delta_t < f_0$ we update the current model time to $t_1=t_0+\Delta_t$ and we proceed in executing the first stochastic event, sample the new waiting time $\Delta_t$ and compare it again with $f_0$.
If otherwise $t_0+\Delta_t>f_0$ we update the current model time to $t_1=f_0$, we proceed in executing the first fixed time event; then we push the next first event (happening time $f_1$) in the stack and compare $t_0 + \Delta_t$.

Let's write down in detail the diagram workflow depicted in Fig.~~\ref{fig:workflow} . W.l.o.g. we start from time $t_{n-1}$:
\begin{enumerate}
    \item Sample $\Delta_t$ from $T\sim \text{exp}(\lambda)$
    \item Compute the new system's time $t_n=t_{n-1}+\Delta_t$
    \item Compare $t_n$ with the time the next fixed time event should happen, defined as $f_m$
    \begin{enumerate}
        \item If $t_n>f_m$ then execute the related fixed time event, increment $m$ by one, and go back to step 3.
        \item If $t_n\leq f_m$ then the random event happens first. Select the related type of random event: it may be an attestation gossip event with probability $\lambda_{\text{attestation}}/(\lambda_{\text{attestation}}+\lambda_{\text{block}})$ or a block gossip event with probability $\lambda_{\text{block}}/(\lambda_{\text{attestation}}+\lambda_{\text{block}})$. Once the event has been selected and executed, go back to step 1.
    \end{enumerate}
\end{enumerate}

\begin{figure}[!t]
\centerline{
    \includegraphics[width=\columnwidth]{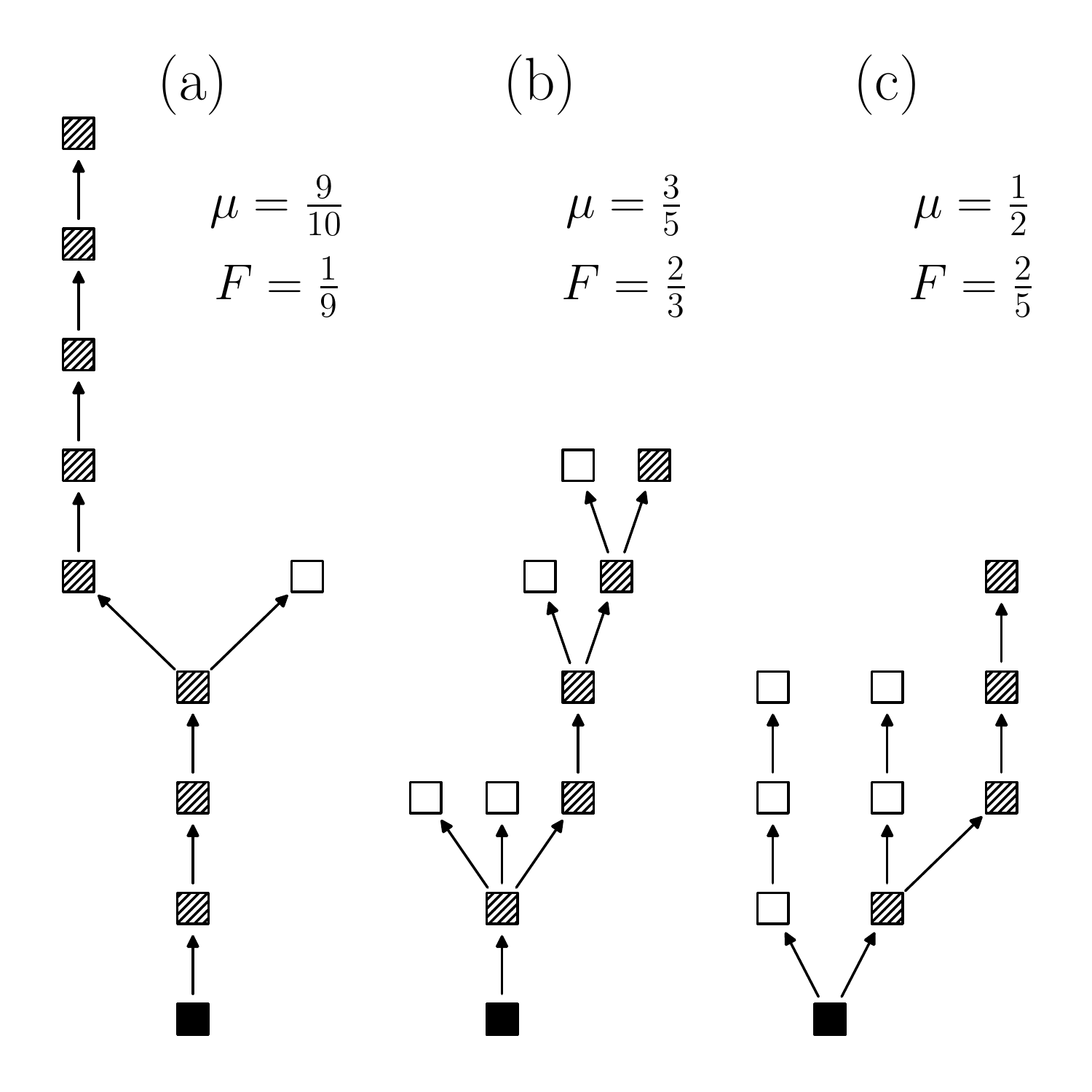}}
    \caption{Illustration of three different blocktree structures, with the genesis block colored in black and the mainchain blocks hatched. In each subplot the mainchain rate ($\mu$) and the branching ratio ($F$) are indicated.}
    \label{fig:blocktree_example}
\end{figure}

\section{Results}\label{sec:results}

In this section we will present the results obtained by simulating our model on multiple realisations and multiple sets of parameters. Here we present two different measures of consensus, the mainchain rate and the branching ratio, which will give us a deeper insight on how the model's parameter affect Ethereum's consensus.
We start by observing that the definition of consensus is not straightforward and unique on a blockchain.
Our goal would be to estimate the rate of agreement between validators in the system, by measuring the number of wasted blocks, meaning blocks not included in the mainchain, the canonical blockchain recognized and accepted by all participants, and the number of forks occurring in the blocktree.
In order to do so we need a unique and clear way to define the mainchain in PoS Ethereum. 
The definition of the mainchain depends on the point of view of each validator in the peer-to-peer network, because the application of the LMD-Ghost rule and consequently the identification of the mainchain strictly depends not only on the observed blocks but also on the observed attestations.
For this reason when we refer to the mainchain in the current context, we will refer to the point of view of an omniscient observer which is always aware of all latest attestation issued and latest blocks gossiped.
While this does not allow us to fully replicate the consensus perception of a validator in the system, it definitely provide a uniquely defined idea of consensus which we can apply and measure. 
Another way to think about the omniscient observer is to think about it, as a node without any latency. In fact, should we have an immediate exchange of information and hence no we witness no latency in the system, each node would be omniscient and thus consensus would be given with only benevolent actors in the system.

Consider $B$ the set of all blocks created during the simulation, $\Theta$ the set of all blocks orphaned during the simulation, and $M$ the set of all blocks in the mainchain(the canonical chain).

\begin{figure*}[!t]
    \centering
    \includegraphics[width=.9\textwidth]{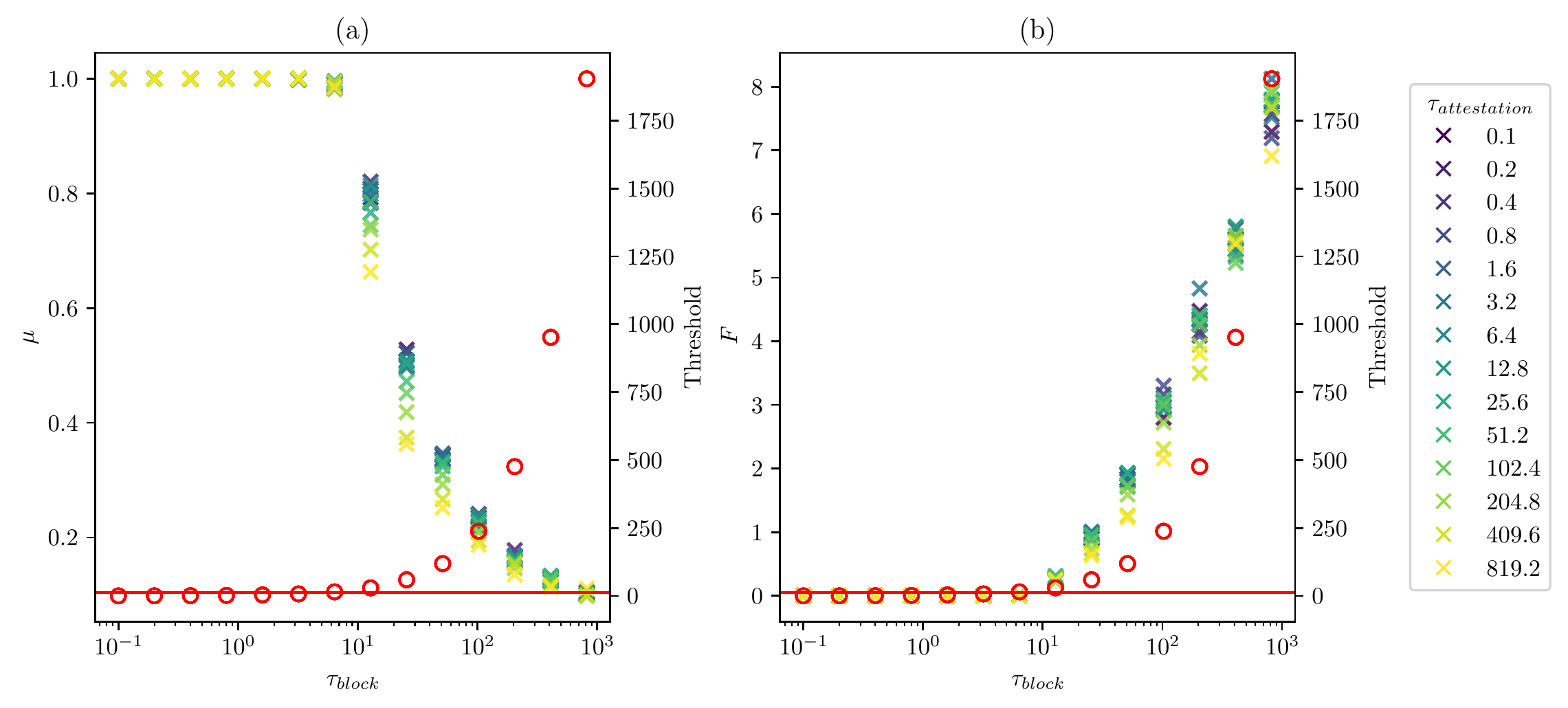}
    \caption{$\mu$ ((a), equation \ref{eq:mainchain_rate}) and $F$ ((b), equation \ref{eq:branch_ratio}) averaged over 20 simulations with varying $\tau_{\text{block}}$ (on the x-axis) and $\tau_{\text{attestation}}$ (in the colour bar). Each simulation is ran over an Erdős–Rényi random graph\cite{erdHos1960evolution} of $N=128$ and average degree $\langle d \rangle=8$. Each simulation runs for $300 s$. The red dots represent the threshold value(defined in equation \ref{eq:threshold}) computed on the ER graph for the specific $\tau_{\text{block}}$. Be aware that the red dots refer to the right axis, while the crosses refer to the left axis, in both panel (a) and panel (b).}
    \label{fig:tau_block}
\end{figure*}

 Define the mainchain rate as the total number of blocks included in the mainchain, denoted by $|M|$, divided by the total number of produced blocks, denoted as $|B|$:
 \begin{equation}\label{eq:mainchain_rate}
      \mu = \frac{|M|}{|B|} = 1 - \frac{|\Theta|}{|B|}
 \end{equation}
 Observe that $\mu \in (0,1]$ and that efficient consensus is realised for $\mu \to 1$, meaning that when all blocks produced are accepted in the mainchain PoS consensus is working smoothly(as opposed to Proof-of-Work consensus where block over-production is a natural feature of the system \cite{nakamoto2008bitcoin}).

Another interesting quantity is the branching ratio, which measures how often forks from the mainchain happen:
\begin{equation}\label{eq:branch_ratio}
    F = \frac{1}{|M|}\sum_{b \in M}\sum_{c \in \Theta} \delta(p(b), p(c))
\end{equation}
where $p(b)$ is the block parent of block $b$ and $\delta$ is Kronecker delta: $\delta(i,j)=1$ if $i=j$ and $0$ otherwise. From the definition it follows that $F\in[0, +\infty)$; for the purpose of our analysis it is important to observe that a perfectly efficient consensus is obtained when $F=0$, meaning when there are no forks.

Fig.~ \ref{fig:blocktree_example} shows an illustrative example of how the mainchain rate and the branching ratio are derived from the structure of the block tree. Panel a of Fig.~ \ref{fig:blocktree_example} shows a (relatively) well formed blockchain. Little orphaned blocks were produced and the block tree exhibits a clear mainchain. Panel b shows an intermediate situation were a mainchain forms but, as the communication in the network is hampered, nodes proposing blocks subsequently may have not yet received their predecessors block. A high branching ratio and an intermediate mainchain result from this behavior. Finally in panel c a situation is considered were latency is very large, such that nodes may not anymore effectively communicate their information. We observe the \emph{privatization} of the block tree, where nodes stay on their respective blockchain and do not anymore work together, leading to a low mainchain rate of block and a low branching ratio.
In this situation the main chain may be ill defined as it will only emerge in this form to the omniscient observer but not the single agent of the system.

Finally, we conducted an experiment on Ethereum PoS consensus tuning the model parameters and observing how $F$ and $\mu$, the consensus proxy variables, are affected in return. 
$\tau_{\text{block}}$ and $\tau_{\text{attestation}}$ vary from $10^{-1}$ to $9\cdot10^{2}$, while the peer-to-peer networks are sampled from an Erdős–Rényi ensemble of $N=128$ nodes and average degree $\langle d \rangle =8$. For each parameter input set we executed $20$ simulations and averaged the results.
As one can see from Fig.~ \ref{fig:tau_block} (a) the mainchain rate decreases as the latency (of blocks) introduced in the system increases. An important observation is the phase transition from the state where all blocks are ordered accordingly to their production and hence forming a straight chain and the point at which forks begin to happen and the blockchain becomes degenerate, i.e. it gets closer to a proper tree rather than a block chain. We notice that the point where this phase transition happens is closely connected to the time between block proposals (i.e. the slot length).
This is natural as when the time between two consecutive block proposals becomes too short, blocks are not able to diffuse to all the network and therefore agents will continue their chain to the best of their knowledge - and form a fork as the block competes with another block at the same height.  
The diameter of the network \cite{chung2001diameter}, denoted by $D(\mathcal{G})$, and the expected time it takes for a block to traverse an edge, $\tau_{\text{block}}$, in an interplay with the time between consecutive block issuance (i.e. the length of a slot) dictate when consensus falls apart in the network. 
The phase transition from an ordered to unordered state takes place when:
\begin{equation}\label{eq:threshold}
D(\mathcal{G}) \, \tau_{\text{block}} > T_{\text{slot}}
\end{equation}
and thus consensus starts to disintegrate.

\begin{figure*}[!t]
    \centering
    \includegraphics[width=.9\textwidth]{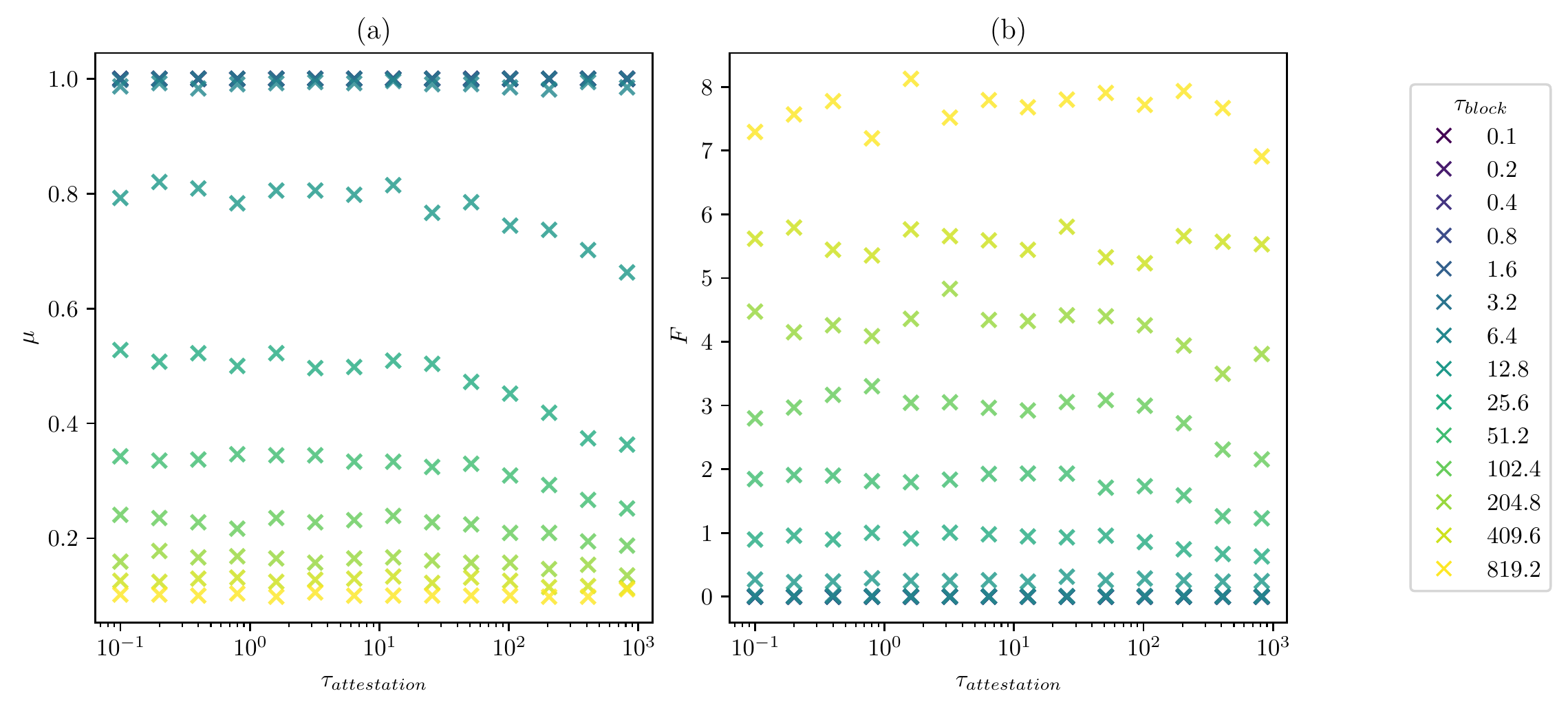}
    \caption{$\mu$ ((a), equation \ref{eq:mainchain_rate}) and $F$ ((b), equation \ref{eq:branch_ratio}) averaged over 20 simulations with varying $\tau_{\text{attestation}}$ (on the x-axis) and $\tau_{\text{block}}$ (in the colour bar). Each simulation is ran over an Erdős–Rényi random graph\cite{erdHos1960evolution} of $N=128$ and average degree $\langle d \rangle=8$. Each simulation runs for $300 s$ (model time).}
    \label{fig:tau_attestation}
\end{figure*}

We observe that while the relation described in equation \ref{eq:threshold} holds for general peer-to-peer networks, we are able to express analytically the Diameter of Erdős–Rényi random graphs for specific parameter choices.
In the specific, we know that for ${1}/{N}<p<\log{N}/{N}$(where $p={\langle d \rangle}/({N-1})$ is the probability of a link forming between two random nodes after ER ensemble) the resulting ER graph will have a giant connected component(meaning there exists a path connecting the quasi-totality of nodes) and the Diameter will concentrate around the value\cite{chung2001diameter}:
\begin{equation}\label{eq:er_diameter}
    D(G^{ER})=\frac{\log{N}}{\log{Np}}.
\end{equation}
Combining together equations \ref{eq:threshold} and \ref{eq:er_diameter} we are able to predict the systems phase transition away from consensus. The results are plotted in red in Fig.~\ref{fig:tau_block} alongside $F$ and $\mu$ values: each dot is $D(\mathcal{G}) \, \tau_{\text{block}}$ value, while the horizontal red line marks the $T_{\text{slot}}$ threshold. As predicted, when the dots overcome the threshold line the phase transition starts.

A solution to increase the robustness of consensus would be to increase the slot length, as blocks have more time to traverse the network, proposers are more likely to know about the whole canonical chain. The obvious downside of such a solution is a decreased output of blocks in a given period. This in turn may decrease the throughput of information in the system, if the block size is not increased. Please note, an increase of the block size may increase the latency in the system. 

Observing the latency for attestations, we see that consensus is mainly driven by the latency inherent to block exchange and the relationship described before. Nonetheless 
we may notice a mediating effect of low attestation latency $\tau_{\text{attestation}}$.

Fig.~\ref{fig:tau_block} panel (b) shows the branching ratio with increasing block latency. We observe that as the systems flow of block information deteriorates the branching ratio increases as expected. In combination with panel (a) we can see how the structure of the resulting block tree mainly depends on the latency of blocks. 

The influence of the attestation latency seems more ambiguous. Fig.~\ref{fig:tau_attestation} exposes its influence on our two measurements and shows the mediating effect of a low attestation latency on consensus in panel (a). Panel (b) suggests that a faster attestation exchange between agents allows them to better agree on a mainchain, even if the latest blocks did not yet arrive. We observe a higher mainchain rate and a slightly increased branching ratio for low values of $\tau_{\text{attestation}}$. This leads us to the conclusion, that nodes may not know about the latest proposed blocks yet, but attestations flow allows them to agree (faster) on a canonical (main-) chain for older blocks. Hence the deviation is more likely to be from the canonical chain and not as a result of a privatization of the blockchain. In other words a lower attestation latency may not necessarily influence how many blocks of a certain \emph{height} are proposed but they may be attached more likely to a part of the main chain which is agreed upon. Coming back to our example in Fig.~\ref{fig:blocktree_example}, one can think about a low attestation latency as a mediating factor, affecting the block tree to resemble the one depicted in panel (b) more than the one depicted in panel (c).

\section{Conclusion}\label{sec:conclusion}
The results discussed in section \ref{sec:results} show us that block gossip latency is the main driver of Ethereum's Proof-of-Stake consensus. Attestation gossip latency has a subordinate role in the formation consensus.
A possible explanation for this phenomenon is that all the validators in the system are assumed to be honest: if we were to relax the honesty assumption and introduce malicious agents in the model, an attacker may take strategic advantage of attestation release in order to split or delay consensus \cite{schwarz-schilling2021attacks, neu2021ebb, neuder2021low}. It must also be noted that in the present work we focused on local consensus rather than finality, the latter may be more sensitive to variations of the attestation gossip latency.

Another interesting result is the detection and prediction of a phase transition out of consensus of the system, depending solely on the block gossip latency, the time between block proposals and the diameter of the underlying Erdős–Rényi peer-to-peer network.

Future directions for the research are the introduction of organized malicious agents executing known attack vectors, the eventual development of new attack strategies based on the attackers knowledge of the peer-to-peer network structure, dynamic validator sets and dynamic stake distributions.

\section*{Acknowledgment}
The authors thank Barnab\'{e} Monnot for the fruitful discussions. 
\bibliographystyle{IEEEtran}
\bibliography{references.bib}

\begin{thebibliography}{10}
\providecommand{\url}[1]{#1}
\csname url@samestyle\endcsname
\providecommand{\newblock}{\relax}
\providecommand{\bibinfo}[2]{#2}
\providecommand{\BIBentrySTDinterwordspacing}{\spaceskip=0pt\relax}
\providecommand{\BIBentryALTinterwordstretchfactor}{4}
\providecommand{\BIBentryALTinterwordspacing}{\spaceskip=\fontdimen2\font plus
\BIBentryALTinterwordstretchfactor\fontdimen3\font minus
  \fontdimen4\font\relax}
\providecommand{\BIBforeignlanguage}[2]{{%
\expandafter\ifx\csname l@#1\endcsname\relax
\typeout{** WARNING: IEEEtran.bst: No hyphenation pattern has been}%
\typeout{** loaded for the language `#1'. Using the pattern for}%
\typeout{** the default language instead.}%
\else
\language=\csname l@#1\endcsname
\fi
#2}}
\providecommand{\BIBdecl}{\relax}
\BIBdecl

\bibitem{Tasca_Tessone_2019}
\BIBentryALTinterwordspacing
P.~Tasca and C.~J. Tessone, ``A taxonomy of blockchain technologies: Principles
  of identification and classification,'' \emph{Ledger}, vol.~4, Feb. 2019.
  [Online]. Available: \url{http://ledger.pitt.edu/ojs/ledger/article/view/140}
\BIBentrySTDinterwordspacing

\bibitem{tree}
F.~Spychiger, P.~Tasca, and C.~J.~Tessone, ``Unveiling the importance and
  evolution of design components through the "tree of blockchain",''
  \emph{Frontiers in Blockchains}, vol.~3, 2021.

\bibitem{bez2019scalability}
M.~Bez, G.~Fornari, and T.~Vardanega, ``The scalability challenge of ethereum:
  An initial quantitative analysis,'' in \emph{2019 IEEE International
  Conference on Service-Oriented System Engineering (SOSE)}.\hskip 1em plus
  0.5em minus 0.4em\relax IEEE, 2019, pp. 167--176.

\bibitem{nakamoto2008bitcoin}
S.~Nakamoto, ``Bitcoin: A peer-to-peer electronic cash system,''
  \emph{Decentralized Business Review}, p. 21260, 2008.

\bibitem{buterin2020combining}
V.~Buterin, D.~Hernandez, T.~Kamphefner, K.~Pham, Z.~Qiao, D.~Ryan, J.~Sin,
  Y.~Wang, and Y.~X. Zhang, ``Combining ghost and casper,'' \emph{arXiv
  preprint arXiv:2003.03052}, 2020.

\bibitem{buterin2017casper}
V.~Buterin and V.~Griffith, ``Casper the friendly finality gadget,''
  \emph{arXiv preprint arXiv:1710.09437}, 2017.

\bibitem{sompolinsky2015secure}
Y.~Sompolinsky and A.~Zohar, ``Secure high-rate transaction processing in
  bitcoin,'' in \emph{International Conference on Financial Cryptography and
  Data Security}.\hskip 1em plus 0.5em minus 0.4em\relax Springer, 2015, pp.
  507--527.

\bibitem{d2022no}
F.~D'Amato, J.~Neu, E.~N. Tas, and D.~Tse, ``No more attacks on proof-of-stake
  ethereum?'' \emph{arXiv preprint arXiv:2209.03255}, 2022.

\bibitem{daian2019flash}
P.~Daian, S.~Goldfeder, T.~Kell, Y.~Li, X.~Zhao, I.~Bentov, L.~Breidenbach, and
  A.~Juels, ``Flash boys 2.0: Frontrunning, transaction reordering, and
  consensus instability in decentralized exchanges,'' \emph{arXiv preprint
  arXiv:1904.05234}, 2019.

\bibitem{neuder2021low}
M.~Neuder, D.~J. Moroz, R.~Rao, and D.~C. Parkes, ``Low-cost attacks on
  ethereum 2.0 by sub-1/3 stakeholders,'' \emph{arXiv preprint
  arXiv:2102.02247}, 2021.

\bibitem{neu2021ebb}
J.~Neu, E.~N. Tas, and D.~Tse, ``Ebb-and-flow protocols: A resolution of the
  availability-finality dilemma,'' in \emph{2021 IEEE Symposium on Security and
  Privacy (SP)}.\hskip 1em plus 0.5em minus 0.4em\relax IEEE, 2021, pp.
  446--465.

\bibitem{schwarz-schilling2021attacks}
\BIBentryALTinterwordspacing
C.~Schwarz{-}Schilling, J.~Neu, B.~Monnot, A.~Asgaonkar, E.~N. Tas, and D.~Tse,
  ``Three attacks on proof-of-stake ethereum,'' \emph{CoRR}, vol.
  abs/2110.10086, 2021. [Online]. Available:
  \url{https://arxiv.org/abs/2110.10086}
\BIBentrySTDinterwordspacing

\bibitem{gillespie1976general}
D.~T. Gillespie, ``A general method for numerically simulating the stochastic
  time evolution of coupled chemical reactions,'' \emph{Journal of
  computational physics}, vol.~22, no.~4, pp. 403--434, 1976.

\bibitem{tessone2021stoc}
C.~J. Tessone, P.~Tasca, and F.~Iannelli, ``Stochastic modelling of blockchain
  consensus,'' \emph{arXiv preprint arXiv:2106.06465}, 2021.

\bibitem{CSS2021}
C.~Schwarz-Schilling, S.-N. Li, and C.~J. Tessone, ``Agent-based modelling of
  strategic behavior in pow protocols,'' in \emph{2021 third international
  conference on blockchain computing and applications (BCCA)}.\hskip 1em plus
  0.5em minus 0.4em\relax IEEE, 2021, pp. 111--118.

\bibitem{CSS2022}
------, ``Stochastic modelling of selfish mining in proof-of-work protocols,''
  \emph{Journal of Cybersecurity and Privacy}, vol.~2, pp. 292--310, 5 2022.

\bibitem{Li2020}
S.-N. Li, Z.~Yang, and C.~J. Tessone, ``Proof-of-work cryptocurrency mining: a
  statistical approach to fairness,'' in \emph{2020 IEEE/CIC international
  conference on communications in China (ICCC workshops)}.\hskip 1em plus 0.5em
  minus 0.4em\relax IEEE, 2020, pp. 156--161.

\bibitem{Li2020b}
------, ``Mining blocks in a row: A statistical study of fairness in bitcoin
  mining,'' in \emph{2020 IEEE International Conference on Blockchain and
  Cryptocurrency (ICBC)}.\hskip 1em plus 0.5em minus 0.4em\relax IEEE, 2020,
  pp. 1--4.

\bibitem{Kraner2022}
B.~Kraner, S.-N. Li, A.~S. Teixeira, and C.~J. Tessone, ``Agent-based modelling
  of bitcoin consensus without block rewards,'' in \emph{2022 IEEE
  International Conference on Blockchain (Blockchain)}.\hskip 1em plus 0.5em
  minus 0.4em\relax IEEE, 2022, pp. 29--36.

\bibitem{fadda2022consensus}
E.~Fadda, J.~He, C.~J. Tessone, and P.~Barucca, ``Consensus formation on
  heterogeneous networks,'' \emph{EPJ Data Science}, vol.~11, no.~1, p.~34,
  2022.

\bibitem{yin2018hotstuff}
M.~Yin, D.~Malkhi, M.~K. Reiter, G.~G. Gueta, and I.~Abraham, ``{HotStuff}:
  {BFT} consensus with linearity and responsiveness,'' in \emph{Symposium on
  Principles of Distributed Computing}, ser. PODC '19.\hskip 1em plus 0.5em
  minus 0.4em\relax ACM, 2019, p. 347–356.

\bibitem{erdHos1960evolution}
P.~Erd{\H{o}}s, A.~R{\'e}nyi \emph{et~al.}, ``On the evolution of random
  graphs,'' \emph{Publ. Math. Inst. Hung. Acad. Sci}, vol.~5, no.~1, pp.
  17--60, 1960.

\bibitem{chung2001diameter}
F.~Chung and L.~Lu, ``The diameter of sparse random graphs,'' \emph{Advances in
  Applied Mathematics}, vol.~26, no.~4, pp. 257--279, 2001.

\end{thebibliography}

\end{document}